\newif\ifabstract
\abstracttrue   
\abstractfalse   

\ifabstract  

\documentstyle[twocolumn]{article}
\makeatletter
\def\section{\@startsection {section}{1}{\z@}{-3.5ex plus -1ex minus
      -.2ex}{2.3ex plus .2ex}{\large\bf}}
\makeatother

\else  

\documentstyle[aps]{revtex}
\begin{document}
\draft
\flushbottom

\fi 

\def\8{\infty}
\def\oh{\frac{1}{2}}
\def\ot{\frac{1}{3}}
\def\oq{\frac{1}{4}}
\def\tt{\frac{2}{3}}
\def\ft{\frac{4}{3}}
\def\tq{\frac{3}{4}}
\def\d{\partial}
\def\i{\imath\,}
\def\ih{\frac{\imath}{2}\,}
\def\undertext#1{\vtop{\hbox{#1}\kern 1pt \hrule}}
\def\ra{\rightarrow}
\def\lfa{\leftarrow}
\def\Ra{\Rightarrow}
\def\lra{\longrightarrow}
\def\ler{\leftrightarrow}
\def\lrb#1{\left(#1\right)}
\def\O#1{O\left(#1\right)}
\def\VEV#1{\left\langle\,#1\,\right\rangle}
\def\tr{\hbox{tr}\,}
\def\trb#1{\tr\lrb{#1}}
\def\dd#1{\frac{d}{d#1}}
\def\dbyd#1#2{\frac{d#1}{d#2}}
\def\pp#1{\frac{\partial}{\partial#1}}
\def\pbyp#1#2{\frac{\partial#1}{\partial#2}}
\def\ff#1{\frac{\delta}{\delta#1}}
\def\fbyf#1#2{\frac{\delta#1}{\delta#2}}
\def\pd#1{\partial_{#1}}
\def\br{\\ \nonumber & &}
\def\brr{\right. \\ \nonumber & &\left.}
\def\inv#1{\frac{1}{#1}}
\def\be{\begin{equation}}
\def\ee{\end{equation}}
\def\bea{\begin{eqnarray} & &}
\def\eea{\end{eqnarray}}
\def\ct#1{\cite{#1}}
\def\rf#1{(\ref{#1})}
\def\EXP#1{\exp\left(#1\right)}
\def\TEXP#1{\hat{T}\exp\left(#1\right)}
\def\INT#1#2{\int_{#1}^{#2}}
\def\MAT{{\it Mathematica }}
\def\LHS{left-hand side }
\def\RHS{right-hand side }
\def\COM#1#2{\left\lbrack #1\,,\,#2\right\rbrack}
\def\AC#1#2{\left\lbrace #1\,,\,#2\right\rbrace}
\def \PDF{probability distribution function }
\def\t{\tilde}
%
\def\pct#1{\input epsf \centerline{ \epsfbox{#1.ps}}}

\author{Victor Gurarie$^{1}$ and Chetan Nayak$^{2}$}

\address{
$^{1}$ Institute for Advanced Study, Olden Lane, Princeton, NJ 08540\\
$^{2}$ Institute for Theoretical Physics, University of California,
Santa Barbara, CA 93106-4030 }

\title{A Plasma Analogy and Berry Matrices for Non-Abelian
Quantum Hall States}

\maketitle

\begin{abstract}

We present an approach to the computation of the non-Abelian
statistics of quasiholes in quantum Hall states, such
as the Pfaffian state, whose
wavefunctions are related to the conformal blocks of
minimal model conformal field theories.
We use the Coulomb gas construction of these
conformal field theories to formulate a plasma analogy
for the quantum Hall states. A number of properties of the
Pfaffian state follow immediately, including the Berry phases
which demonstrate the quasiholes' fractional charge,
the abelian statistics of the two-quasihole state,
and equal-time ground state correlation functions.
The non-Abelian statistics of multi-quasihole states
follows from an additional assumption. 

\end{abstract}
\vspace{3mm}

{\it Introduction.} Recent investigations of the possibility of finding
non-Abelian braiding statistics in the quantum Hall regime
have used a remarkable observation due to Moore and Read as a springboard
\cite{mooreread,blokwen,wenwu,wilczek,gfn}.
They drew attention to the fact that certain wavefunctions
which have been proposed to describe incompressible quantum Hall states
could be represented as conformal blocks in corresponding
conformal field theories. This is not a statement about
dynamics in the quantum Hall regime\footnote{At least
on the face of it. In fact, in many cases
the conformal field theory which
reproduces the wavefunctions is the same
as the one which describes
the gapless excitations at the edge.
If the effective field theory
of the bulk state is a Chern-Simons theory,
this relationship can be understood in light of
Witten's discoveries \cite{witten}:
the states of the Chern-Simons
theory which describes the bulk are isomorphic to
the conformal blocks of the CFT which
reproduces the bulk wavefunction.
At the same time, the boundary theory
which is concomitant with the Chern-Simons theory is precisely
the CFT of the edge. The equality of these CFTs follows
from the general covariance of the Chern-Simons theory.}, but
merely a statement that some wavefunctions could be
reconstructed from conformal
field theories. Hence, it might not appear particularly
useful. However, this observation suggests a way
around a technical impasse which impedes the
study of non-Abelian quantum Hall states.
The problem and a strategy for its circumvention noted
by Moore and Read may be summarized as follows.
Suppose we initially eschew
any attempt to realistically account for the physics of the long-range
Coulomb interaction, finite thickness of the quantum well,
inter-Landau-level mixing, and other complications, and, instead,
work with an unrealistic but tractable model Hamiltonian
whose exact zero-energy eigenstates
(ground state and multi-quasihole states) can be
explicitly constructed. The statistics of these quasiholes
may be determined from the exact eigenstates using the Berry
phase technique \cite{asw}. If the model Hamiltonian may be
adiabatically connected to the one which
governs the actual physical 2D electron gas -- i.e. if
the two Hamiltonians are in the same universality class --
the statistics of the quasiholes and other `topological
quantum numbers' \cite{topqn} will be the same for both.
In such a case, the calculation of these properties of the idealized
Hamiltonian will be relevant to the real system.
The stumbling block
is that Berry phases (or Berry matrices in the non-Abelian case)
are determined from matrix elements
of the eigenstates and further progress is impossible
unless we can find a method for calculating the
matrix elements of the correlated many-body
wavefunctions of interest.

In his pioneering work on the
fractional quantum Hall effect \cite{laughlin},
Laughlin took advantage of the equality between the squared modulus
of his wavefunctions and the Boltzmann weight of a classical
two-dimensional plasma. The conventional wisdom on plasmas
implies that the Laughlin states are quantum liquids \cite{laughlin}
and has also facilitated calculations of off-diagonal long-range-order
in these states \cite{topqn}. The original calculation of the
fractional statistics of the Laughlin quasiholes \cite{asw}
relied on the incompressible liquid nature
of the states but did not explicitly utilize the plasma analogy.
However, the calculation may be rephrased \cite{blokwen}
in the following way so that its dependence on the plasma analogy
is emphasized. Let us write the wavefunction for a Laughlin
state with two quasiholes at $\eta_1$ and $\eta_2$
in the following form ($\ell_0$ is the magnetic length):
\be
{\Psi_{{\eta_1}{\eta_2}}}\, = \,
{({\eta_1}-{\eta_2})^{1/m}}
\,\,\,{\prod_k}\,({z_k}-{\eta_1})\,({z_k}-{\eta_2})\,\,\,
{\prod_{i>j}}\, {({z_i} - {z_j})^m}\,\,
{e^{-\frac{1}{4{\ell}_0^2}\,
\frac{1}{m}\left({|{\eta_1}|^2}+{|{\eta_2}|^2}\right)}}
\,\,{e^{-\frac{1}{4{\ell}_0^2} \sum |z_i|^2 }}~.
\label{fracstatlaugh}
\ee
Ordinarily, we would choose the wavefunction to be a single-valued
function of the quasihole coordinates (which are simply some
parameters in an electron wavefunction) by, for instance,
replacing the first factor by its modulus or removing it altogether.
In (\ref{fracstatlaugh}),
we have intentionally chosen the phase of the wavefunction
such that it is a multi-valued function of the quasihole coordinates
(but, of course, single-valued in the electron coordinates);
the net effect of braiding the two quasiholes will
be the Berry holonomy together with the explicit monodromy of
(\ref{fracstatlaugh}). Only this combination is physically observable,
and by changing the phase of the wavefunction -- i.e. performing
a gauge transformation -- we can divide
this phase between the Berry holonomy and the explicit monodromy
in any way we wish. As we will see momentarily, the choice
(\ref{fracstatlaugh}) is particularly convenient because it is
the `fractional statistics gauge' \cite{frank} in which the Berry holonomy
vanishes and the explicit monodromy is the whole story. A
second advantage of (\ref{fracstatlaugh}) is that it
is a holomorphic function of ${\eta_1},{\eta_2}$ up to
the Gaussian factor. Hence, the Berry holonomy, $\gamma$, acquired when
$\eta_1$ is taken adiabatically around $\eta_2$ is:
\be
\gamma = \oint d{\eta_1}\, \langle {\Psi_{{\eta_1}{\eta_2}}}|\,
\frac{\partial}{\partial{\eta_1}} |{\Psi_{{\eta_1}{\eta_2}}}\rangle
=  \oint d{\eta_1}\,\frac{\partial}{\partial{\eta_1}}
\, \langle {\Psi_{{\eta_1}{\eta_2}}}|{\Psi_{{\eta_1}{\eta_2}}}\rangle
+ \frac{\pi}{m}\,\frac{\Phi_{ }}{\Phi_0}
\ee
where $\Phi$ is the magnetic flux enclosed within the
path taken by $\eta_1$ and $\Phi_0$ is the flux quantum.
The second equal sign follows from the fact that
$\langle {\Psi_{{\eta_1}{\eta_2}}}|$ depends only on
${\bar \eta}_1$ and not on $\eta_1$, except through the
Gaussian factor, which yields the second term on the
right hand side. This term represents the fractional charge
of the quasiparticles and does not depend on whether or not
$\eta_1$ encircles $\eta_2$. We now argue that as a
result of the prefactor chosen in (\ref{fracstatlaugh}),
the first term vanishes. This follows from the
observation that this inner product
is the partition function of a classical plasma of
charge $\sqrt{m}$ particles at temperature $T=1$
with a neutralizing background
of density $1/2\pi\sqrt{m}$ and two fixed charge
$1/\sqrt{m}$ particles at $\eta_1$ and $\eta_2$:
\be
\langle {\Psi_{{\eta_1}{\eta_2}}}|{\Psi_{{\eta_1}{\eta_2}}}\rangle =
\int\,{\prod_i}{d^2}{z_i}\,{e^{ \frac{2}{m}\ln|{\eta_1}-{\eta_2}|
+ {\sum_{i,\alpha}} 2\ln|{\eta_\alpha}-{z_j}|
+ {\sum_{i>j}} 2m\ln|{z_i}-{z_j}|
- \frac{1}{m}\,\frac{1}{4{\ell}_0^2}\left({|{\eta_1}|^2}+{|{\eta_2}|^2}\right)
- \frac{1}{4{\ell}_0^2} \sum |z_i|^2  }}
\label{laughplasma}
\ee
(Usually, one takes $T=m$ and has charge $m$ electrons
and charge $1$ quasiholes, but the above convention
is more natural for what follows.)
When $|{\eta_1}-{\eta_2}|$ is larger
than the Debye screening length,
the interaction between these two charges will
be screened, so
the partition function will
be independent of $\eta_1$, $\eta_2$. In other words,
$\langle {\Psi_{{\eta_1}{\eta_2}}}|{\Psi_{{\eta_1}{\eta_2}}}\rangle$
is independent of $\eta_1$ and $\eta_2$ so long as they are
well-separated. Therefore, the
Berry holonomy vanishes and the fractional statistics is given by
the explicit monodromy of (\ref{fracstatlaugh}).

Unfortunately, this argument appears limited to the
Laughlin states; the squared moduli of the states
which are promising hunting
grounds for non-Abelian statistics, such as the Pfaffian
state,
\bea
{\Psi_{\rm Pf}}\,\, =\,\,
{\rm Pf}\left(\frac{1}{{z_i} - {z_j}}\right)
\,\,\,{\prod_{i>j}}{\left({z_i} - {z_j}\right)^2}\,\,\,
{e^{-\frac{1}{4{\ell}_0^2} \sum |z_i|^2 }}
\label{pfstate}
\eea
(where ${\rm Pf}\left(\frac{1}{{z_i} - {z_j}}\right) =
{\cal A}\left(\frac{1}{{z_1} - {z_2}}\,
\frac{1}{{z_3} - {z_4}}\,\ldots\right)$ is the antisymmetrized product
over pairs of electrons)
do not seem to be equal to the Boltzmann weight of a classical
plasma, at least not in the most obvious way.
A further complication arises from the fact that
there is a degenerate set of multi-quasihole states in
the putative non-Abelian cases, so off-diagonal
matrix elements must also be calculated in order to
obtain a Berry matrix. It is at this seemingly hopeless
point that Moore and Read's observation can
come to the rescue. Recall that the conformal blocks of
certain conformal field theories are related to
many of the quantum Hall
states -- including the Laughlin states and
the Pfaffian state -- which are exact ground states of
model Hamiltonians. It is natural to conjecture
that the quasihole statistics
are equal to the monodromy properties of
the conformal blocks (see also \cite{blokwen}). These
statistics are encapsulated in the concomitant
Chern-Simons theories \cite{witten} which describe the long-wavelength
physics of the quantum Hall state \cite{topqn}.
Some justification for this conjecture, on rather general grounds,
was given in \cite{wilczek}, where it was used to conclude that
the braiding matrices of the $2n$-quasihole states in the
Pfaffian are embedded in the spinor representation of
$SO(2n)$. 

In this paper, we give further arguments (which we believe can
serve as an outline for a proof) supporting this conjecture.
We formulate a plasma analogy for non-Abelian quantum
Hall states which makes use of the fortunate fact that
all of the minimal model conformal field theories
have a Coulomb gas description. At a first glance,
this seems tantalizing but insufficient for our purposes
because the construction involves complicated contour integrals
of the locations of screening charges.
However, following Mathur \cite{mathur}, we 
rewrite these contour
integrals as two-dimensional integrals.
As a result,
we find that sums of the desired inner products
can be expressed as the partition functions of
plasmas with an additional large number of screening charges
(which only enhance the screening property which we invoke).
The individual inner products are obtained by analytic continuation.
This approach is applied to the Pfaffian state, which
is associated with the $c = \frac{1}{2} + 1$ conformal
field theory. Our plasma analogy enables us to compute the electron 
density matrix, the charge of the quasihole
excitations  (which, we argue, have localized charge
distributions), and the Abelian statistics of
the two-quasihole state. With one assumption regarding the
analytic continuation,
we can show that the Berry matrices of states with
an arbitrary number of quasiholes are trivial
and the non-Abelian statistics can be obtained entirely
from the monodromy properties of the conformal blocks,
in agreement with \cite{wilczek}.

{\it Background Charge Construction and a Plasma Analogy.} 
Let us momentarily
take one step back before taking two
steps forward. Recall that the Laughlin state
can be expressed as a conformal block of the $c=1$ theory,
which is the theory of a free bosonic field $\varphi$:
\be
S = \frac{1}{2\pi} \int \partial \varphi \bar \partial \varphi
\ee
The Laughlin state is given by the conformal block
of the vertex operator
${V_{\sqrt{m}}}(z) = :\exp (i \sqrt{m} {\varphi_R}(z)):$,
(${\varphi_R}$ is the holomorphic part of the boson field)
which plays the role of the electron, together with the
operator for a neutralizing background charge.
\be
{\Psi_{1/m}} = \langle\, {e^{i\sqrt{m}\phi({z_1})}}\,
{e^{i\sqrt{m}\phi({z_2})}}\,\ldots\,{e^{i\sqrt{m}\phi({z_N})}}
\,{e^{-i \,\int{d^2}z\,\sqrt{m}{\rho_0}\phi(z)}}
\rangle
\label{cblgs}
\ee
Conformal blocks with ${V_{1/\sqrt{m}}}({\eta_i}) =
:\exp (i {\varphi_R}({\eta_i})/\sqrt{m}):$
are equal to states with quasiholes at $\eta_i$.
The product of the Laughlin wave function and its complex conjugate
can be written as the product of a holomorphic and
an antiholomorphic conformal block or, simply, as a
correlation function of the left-right symmetric field
${V_{\sqrt{m}}}(z,{\bar z}) =
:\exp (i \sqrt{m} ({\varphi_R}(z)+{\varphi_L}({\bar z}))):$.
These correlation functions admit the Coulomb gas or
plasma description of (\ref{laughplasma}).
The Pfaffian state, however, involves in its description
a $c={1 \over 2}$ conformal field theory. Can we
describe it, too, by means of a Coulomb gas?

Naively, the answer is yes since 
any conformal field theory admits a Coulomb gas description,
as was discovered by
Feigin and Fuchs and Dotsenko
and Fateev (see \cite{dotfat} and references therein). 
The Dotsenko-Fateev picture involves putting a certain
{\it background charge} at infinity and introducing 
{\it screening operators} in a $c=1$ conformal block. 
Then any conformal block of any conformal field theory
can be expressed as
\be
\label{screen}
F(z_1, z_2, \dots) =
{\int_{{\cal C}_i}} \prod_i d w_i
\left\langle \prod_j O_j (w_j) \prod_k V_k (z_k)
\right\rangle
\ee
where the vertex operators $O({w_j})$ are the screening operators and
${V_k}({z_k})$ are the vertex operators of interest. The integration
of the screening operator locations is along certain contours,
${\cal C}_i$. The choice of contours determines which 
conformal blocks we obtain. Notice that the block 
on the right hand side of \rf{screen} is a standard Coulomb gas.
Therefore it's tempting to conclude that we have a
plasma analogy for any wave
function connected to conformal field theory.

However, this conclusion is premature. The contours for the
integrals of \rf{screen} are in general hard to define. To compute
inner products we would have to introduce holomorphic and
antiholomorphic screening charges, each integrated over
its own contour. Since, in general, the number of
screening charges we need will be of the order of the
number of the electrons, it is unclear
how we can compute the desired inner products. It is also
unclear if we are allowed to exchange the order of integration
over the positions $z_i$ of the electrons (which we will eventually
have to do in order to compute the
inner products) with the contour integrations over the
screening charge locations since these contours
are supposed to wind around
the electron positions in a complicated way \cite{dotfat}.

A way out of this quagmire can be found in a
paper by Mathur \cite{mathur}. He suggested the following.
If, instead of considering holomorphic
vertex operators, we consider full vertex operators which
are the products of the holomorphic and antiholomorphic ones,
$V(z, \bar z) \equiv V(z) V(\bar z)$, then it can be shown that 
\be
\label{screenm} 
\int \prod_i d^2 w_i \left\langle \prod_j O(w_j, \bar w_j) 
\prod_k V(z_k, \bar z_k) \right\rangle = \sum_n C_n F_n(z_1, z_2, \dots)
F_n(\bar z_1, \bar z_2, \dots)
\ee
where the sum over $n$ is the sum over different conformal
blocks $F_n$ (different choices of contours in \rf{screen}) while
the $C_n$ are the so-called {\it structure coefficients}, which
make the combination on the right-hand-side
of \rf{screenm} a single valued function. 

The advantage of \rf{screenm} over \rf{screen} is that the
integration of the screening charges is over the entire
complex plane, thereby putting the screening charges and
electrons on equal footing in the computation of inner products. 
The disadvantage of \rf{screenm} is that
instead of producing the squared modulus of a
particular conformal block (corresponding
to a particular wave function), it gives us a linear 
combination of them. We will discuss ways of getting around this
problem later. For now, let us concentrate on cases in
which there is only one conformal block,
making the summation over $n$ in \rf{screenm} unnecessary.

As an example, we consider the Pfaffian state \rf{pfstate},
which is a conformal block of the $c = \frac{1}{2} + 1$
theory \cite{mooreread,wilczek}.
The Pfaffian factor is the $c={1 \over 2}$ part, 
while the rest comes from the $c=1$ part and has
the standard plasma form of a Laughlin state.
The Laughlin plasma consists of charge $\sqrt{2}$ electrons.
According to \cite{dotfat}, the $c={1 \over 2}$ 
part can also be described
as a plasma with  $N$ charges of magnitude
$3/\sqrt{6}$ or $-2/\sqrt{6}$
associated with the electrons together with $\approx {N \over 2}$
screening charges of magnitude $4/\sqrt{6}$ or $-3/\sqrt{6}$ 
which also roam the entire complex plan and the background
charge $-1/\sqrt{6}$ which has to be taken to the infinitely far
away point. In other words, the background charge, being at
infinity, does not interact with other charges, but shifts
the tolal charge balance. The sum of all the other charges 
of our plasma has to be equal to minus the background charge (details
can be found in \ct{dotfat}). We note that at large $N$ it does not
affect the neutrality of the plasma. 

We note that each operator can be represented by either of two
different charges; this is a characteristic feature of the
background charge construction. 
We have a certain freedom in how we
pick either of these charges for
each operator, and there is more than
one equivalent representation of each conformal block.
One particular construction is discussed in \ct{felder}. However,
these details are not relevant at the moment. For our purposes,
it is sufficient to know that, together with the
background charge, our plasma is neutral,
and that, roughly speaking, we need one screening charge for each
singularity encountered in the wave function.
There are $N \over 2$ singularities in the $c={1 \over 2}$ part of
\rf{pfstate}, making the total
number of screening charges $\propto {N \over 2}$, a number which
is large when the total number of the electrons $N$ is large. 

Therefore we arrive at the central point of this
paper: the Pfaffian state can be mapped to a plasma
{\it with three species of charges which interact via two different
logarithmic interactions}, i.e.
\be
{\left|{\Psi_{\rm Pf}}\right|^2} = \int{\prod_A}{d^2}{w_A}\,\,
{e^{-\beta\Phi({z_i},{\eta_\alpha},{w_A})}}
\label{partialpart}
\ee
The inverse temperature, $\beta$, is $1$, and
$\Phi$ is the potential energy of a classical two-dimensional
gas of logarithmically-interacting particles corresponding
to:

 ({\it a}) electrons, which have charge $\sqrt{2}$
with respect to the first interaction
($c=1$) and charge $3/\sqrt{6}$ or $-2/\sqrt{6}$
with respect to the second ($c=1/2$)

({\it b}) quasiholes, which
have charge $1/2\sqrt{2}$ with respect to the first
interaction ($c=1$) and charge $3/(2 \sqrt{6})$ or 
$-1/2 \sqrt{6}$ with respect to the second ($c=1/2$)

({\it c}) the screening charges, which are neutral with respect to the first
interaction ($c=1$) and carry charge $4/\sqrt{6}$ or $-3/\sqrt{6}$
with respect to the other ($c=1/2$)

({\it d}) the background charge, which is also neutral with
respect to the first interaction ($c=1$) and carry charge
$-1/\sqrt{6}$ with respect to the second ($c=1/2$). 

The squared norm of the Pfaffian state
is the partition function of
this plasma. The squared modulus of the Pfaffian
wavefunction \rf{partialpart} -- which, naively, is a mess --
is, in fact, the Boltzmann weight of this plasma
with the screening charges already integrated
out. The integration
of the $N/2$ $w_A$'s does not affect the $c=1$ interaction and,
at least partially, screens the $c=1/2$ interactions 
among the $z_i$'s and $\eta_\alpha$'s. Hence, if we write
\rf{partialpart} as the Boltzmann weight of some
effective Hamiltonian, $\tilde \Phi$, it will have two
species of particles which interact logarithmically
through the $c=1$ interaction and also through
some weaker interaction (perhaps short-ranged)
which is the partially screened $c=1/2$ interaction.
Though this latter interaction may be very complicated,
we expect that \rf{partialpart} still essentially describes
a plasma even after the $w_A$'s have been integrated out.

{\it Quantum Liquid Nature and Electron Density Matrix.} This observation
allows us to immediately compute various properties
of this state. Following Laughlin, we anticipate that
it is a liquid because the associated plasma is above the crystallization
temperature. As a result, we can compute its electron density matrix
\be
\rho(z, z') = \int \prod_{i=2}^N d^2 z_i \Psi_{Pf}^{*} (z', z_2, \dots) 
\Psi_{Pf}
(z, z_2, \dots)
\ee
We note that $\rho(z,z)$ is given by a partition function of our plasma
with one charge held fixed at $z$; in a liquid, it must be independent of $z$.
On the other hand,
\be
\rho(z,z') = \exp \left(-{1 \over 4} ( |z|^2 + |z'|^2) \right) G(z, \bar z')
\ee
where $G(z, \bar z')$ is analytic in its arguments. The only way
to satisfy these two constraints is if
\be
\rho(z,z') = 
\exp \left( -{1 \over 4} (|z|^2 + |z'|^2) + {1 \over 2} z \bar z' 
\right)
\ee
just as in the Laughlin states. There is no sign of
pairing or any delicate power-law correlations in
the electron Green function. Indeed, we don't expect
such correlations in multi-electron Green functions
either; all of these should
be Gaussian or exponentially localized as a result of the gap.
The interesting correlations should appear in Green
functions of the non-local order parameter \cite{mooreread}.
These correlations are related to the `topological
ordering' of the state and, hence, should be more stable
against local perturbations than putative subtle
correlations in the electron Green functions would
be. Unfortunately the plasma analogy  in the form
presented here does not seem to allow to compute them. 

Finally, we can argue, following Laughlin \cite{laughlin},
that the charges of the fractionally charged particles are
localized around their positions, due to the effective plasma
repulsion. The quasiholes do not have power-law
tails in their charge distribution as one might
have naively guessed from the form of the wavefunctions
and from the fact that they carry half of a flux
quantum.

{\it Fractional Charge and Abelian Statistics
in the Two-Quasihole State.}
There is a unique two-quasihole state (see \cite{wilczek})
if the quasiholes are fixed at positions $\eta_1$
and $\eta_2$. It equal to a conformal block with
twist fields and appropriate $c=1$ vertex operators
inserted to represent the quasiholes \cite{mooreread,wilczek}.
According to our plasma analogy, the squared norm of
the two-quasihole state is the partition function of
our two-component plasma with two external charges
(of magnitudes $1/2\sqrt{2}$ with respect to the $c=1$
interaction and $3/(2 \sqrt{6})$ or  $-1/2 \sqrt{6}$
with respect to the $c=1/2$ interaction)
held fixed at $\eta_1$ and $\eta_2$. The plasma will screen
these charges, so the partition function will be independent
of $\eta_1$ and $\eta_2$ when their separation is larger
than the screening length:
\be
\VEV{\Psi_{Pf}^{\rm 2-holes}
(\eta_1, \eta_2)|\Psi_{Pf}^{\rm 2-holes}(\eta_1,\eta_2)} = 
{Z_{\rm two\, comp.\, plasma}}(\eta_1, \eta_2) = {\rm const.}
\ee
and consequently the non-trivial part of the
Berry connection of these wave functions vanishes
(the trivial part, coming from the Gaussian, gives the fractional charge
of the quasiholes, $1/4$)
\bea
\VEV{  \Psi_{Pf}^{\rm 2-holes}
(\eta_1, \eta_2)\left|{\partial \over \partial \eta_1}\right|
\Psi_{Pf}^{\rm 2-holes}(\eta_1,\eta_2) }\, -\,
{\rm term\,coming\,from\,the\,Gaussian} =\br
{\partial \over \partial \eta_1} \VEV{\Psi_{Pf}^{\rm 2-holes}
(\eta_1, \eta_2)|\Psi_{Pf}^{\rm 2-holes}(\eta_1,\eta_2)} = 0
\eea
Therefore we can read the fractional statistics of the
quasiholes of the Pfaffian state from the monodromy
of the conformal block which, is this case,
is determined by a factor of $(\eta_1-\eta_2)^\alpha$. Interestingly,
$\alpha=0$ or, in other words, in a two-quasihole state,
the two quasiholes are bosonic with respect to each other
\cite{mooreread,wilczek}.
The fractional statistics which is concomitant with
the fractional charge is cancelled by the fractional statistics
resulting from the Pfaffian part of the wavefunction.
Since the charge and statistics
of the quasiholes are different, it should actually be easier
to experimentally disentangle them in the Pfaffian state
than in a Laughlin state.

{\it Non-Abelian Statistics of $2n$-Quasihole States.}
Now let us take up the less trivial case of four quasiholes.
There are two possible four-quasihole wave functions
when the quasiholes are located at $\eta_1$, $\eta_2$, $\eta_3$, $\eta_4$.
We call them $\Psi_0$ and $\Psi_{1 \over 2}$, following \cite{wilczek}.
According to \rf{screenm}, the plasma analogy allows us to compute
only the combination
\be
\VEV{ \Psi_0|\Psi_0} + \VEV{\Psi_{1 \over 2} | \Psi_{1 \over 2}} = 
{Z_{\rm two\, comp.\, plasma}}({\eta_1},{\eta_2},{\eta_3},{\eta_4}) = {\rm const.}
\label{pfaffplasms}
\ee 
This is not enough to argue that the braiding statistics
of the quasiholes can be read off the conformal blocks. 
What we need is the somewhat stronger statement,
\be
\label{inner}
\VEV{\Psi_i | \Psi_j} =  C_{ij}, \ \, i,j =0, {1 \over 2}
\label{desres}
\ee
where $C_{ij}$ is some constant (independent of the positions of quasiholes)
matrix, which, if true, implies that the nonabelian
Berry connection vanishes
\be
\label{Conn}
A_{ij} = \VEV{\Psi_i \left| {\partial \over \partial \eta}\right| \Psi_j} =
{\partial \over \partial \eta} C_{ij} = 0
\ee

To derive \rf{inner} we apparently need to generalize \rf{screenm}.
However, we can instead look at the problem in the following way.
\rf{screenm} represents a single valued function, understood as a function
of $\eta_i$ and ${\bar\eta}_i = \eta_i^*$.
However, as a function of $\eta_i$ only, with ${\bar\eta}_i$
viewed formally as an independent variable,
\rf{screenm} is not single-valued. Moreover, by taking one
$\eta_i$ around another (while keeping ${\bar\eta}_i$ at bay)
we can transform \rf{screenm} into an arbitrary combination of
conformal blocks.
For example, if we take $\eta_1$ around $\eta_2$ while holding
${\bar\eta}_1$ and ${\bar\eta}_2$ as well as $\eta_3$, ${\bar\eta}_3$,
$\eta_4$, ${\bar\eta}_4$ fixed, then ${\Psi_0}\rightarrow {\Psi_0}$
while ${\Psi_{1\over 2}}\rightarrow -{\Psi_{1\over 2}}$. If
we instead take $\eta_1$ around $\eta_4$, while holding
the rest fixed, ${\Psi_0}\rightarrow {\Psi_{1\over 2}}$
and ${\Psi_{1\over 2}}\rightarrow{\Psi_0}$. Hence, with these two
braiding operations we can transform
$\VEV{ \Psi_0|\Psi_0} + \VEV{\Psi_{1 \over 2} | \Psi_{1 \over 2}}$
into $\VEV{ \Psi_0|\Psi_0} - \VEV{\Psi_{1 \over 2} | \Psi_{1 \over 2} }$,
and $\VEV{ \Psi_0|\Psi_{1 \over 2}} \pm \VEV{\Psi_{1 \over 2} | \Psi_0}$,
from which we can obtain the desired inner products \rf{inner}.
In other words, after performing
these braiding operations, we can express
the above combinations of inner products
as analytic continuations of
${Z_{\rm two\, comp.\, plasma}}({\eta_1},{\eta_2},{\eta_3},{\eta_4})$.
The crux of the matter is now: can these analytically
continued partition functions also be interpreted as
partition functions of plasmas, albeit with modified
interactions? If so, then we can, as usual, invoke plasma
screening and conclude that all of the inner
products are independent of the positions of the quasiholes.

At the moment, we do not have a proof that this is so.
However, the following line of reasoning strongly supports
this contention. Earlier, we argued that integration
of the $w_A$'s should result in some sort of plasma with
partially screened interactions.
If we then proceed and integrate out the $z_i$'s, we
expect total screening of the interactions between the
$\eta_\alpha$'s and, hence, the result \rf{pfaffplasms}.
Suppose, instead, that before integrating the $z_i$'s
we first carry out one of the analytic
continuations described above. This will not effect the
$c=1$ interaction since the area enclosed by the path of
the $\eta_\alpha$'s can be made arbitrarily small and
the $c=1$ interaction will only be sensitive to
the analytic continuation if one of the $z_i$'s crosses
this path. The $c=1/2$ interaction can be affected,
but only through the introduction of phases which
can change as $z_i$'s approach $\eta_\alpha$'s.
If these phases interfere destructively, the off-diagonal
matrix elements vanish and the diagonal
matrix elements are equal and constant.
At worst, they can interfere constructively,
in which case the other matrix elements are
in the same situation as \rf{pfaffplasms}.
This scenario, too, leads to the desired result \rf{desres}
if we again appeal to plasma screening.

If this argument proves to be correct, we can conclude (by
straightforwardly extending it to the $2n$-quasihole case)
that the statistics of quasiholes in the Pfaffian state is
given by spinor representations of $SO(2n)$, as was conjectured
in \cite{wilczek}.

{\it Discussion.}
As we discussed in the introduction,
some quantum Hall wavefunctions can be reproduced
by conformal field theories, and it is natural
to conjecture that the braiding properties of the
former are given by those of the latter \cite{mooreread,blokwen}.
In this paper, we have attempted to answer the question `why?'
The basic reason, we submit, is that these
states are all related -- through the background
charge construction \cite{dotfat} -- to Coulomb
gases. As a result of plasma screening in these Coulomb gases,
wavefunctions given by conformal blocks
have trivial Berry connections; their braiding properties
are determined entirely by the monodromies of the
conformal blocks.

Thus far, we have focussed on the Pfaffian state. However, it
is clear from our discussion that the arguments we have used are
quite general and may be applied to any quantum Hall state
with wavefunctions given by the conformal blocks of
a theory with a background charge
construction. To the casual observer, the background
charge representation of a conformal block may
seem like a complicated integral represention.
However, it is perfectly tailored for
the present context because the electron
coordinates must also be integrated in the calculation
of Berry matrices. Consequently, the screening charges
are on the same footing as the electrons and the representation
(\ref{partialpart}) is a natural generalization of
the Coulomb gas representation of the Laughlin states.

\begin{acknowledgements}
We would like to thank T. Spencer, E. D'Hoker, and N. Read
for helpful discussions.
\end{acknowledgements}


\begin{thebibliography}{99}
\bibitem{mooreread} G. Moore and N. Read, {\it Nucl. Phys.}
{\bf B360} (1991) 362.
\bibitem{blokwen} B. Blok and X.G. Wen, {\it Nucl. Phys.}
{\bf B374} (1992) 615.
\bibitem{wenwu} X.G. Wen, Y.S. Wu, {\it Nucl. Phys.}
{\bf B419} (1994) 455.
\bibitem{wilczek} C. Nayak and F. Wilczek, {\it Nucl. Phys.}
{\bf B479} (1996) 529.
\bibitem{gfn} V.Gurarie, M. Flohr, C. Nayak, IASSNS-HEP-97/5,
NSF-ITP-97-014, cond-mat/9701212.
\bibitem{asw} D.P. Arovas, J.R. Schrieffer, and F. Wilczek,
{\it Phys. Rev. Lett.} {\bf 53} (1984) 722.
\bibitem{topqn} S.M. Girvin and A.H. MacDonald,
{\it Phys. Rev. Lett.} {\bf 58} (1987) 1252.
S.C. Zhang, T.H. Hansson, and S. Kivelson, {\it Phys. Rev. Lett.}
{\bf 62} (1988) 82. N. Read, {\it Phys. Rev. Lett.}
{\bf 62} (1988) 86.
X.G. Wen, {\it Int. J. Mod. Phys.}
{\bf B2} (1990) 239; {\it Phys. Rev.} {\bf B40} (1989) 7387.
X.G. Wen and Q. Niu, {\it Phys. Rev.} {\bf B41} (1990) 9377.
\bibitem{laughlin} R.B. Laughlin, {\it Phys. Rev. Lett.} {\bf 50}
(1983)1395 .
\bibitem{frank} F. Wilczek, {\it Phys. Rev. Lett.} {\bf 49}
(1982) 957.
\bibitem{witten} E. Witten, {\it Comm. Math. Phys.} {\bf 121}
(1989) 351.
\bibitem{mathur} S. Mathur, {\it Nucl. Phys.} {\bf B369}
(1992) 433.
\bibitem{dotfat} Vl. S. Dotsenko and V.A. Fateev, {\it Nucl. Phys.} {\bf B240}
(1984) 312; {\bf B251} (1985) 3691.
\bibitem{felder} G. Felder, {\it Nucl. Phys.} {\bf B317} (1989) 215
\end{thebibliography}
\end{document}